%% file: ICASSP19.tex
\documentclass{article}
\usepackage{spconf,amsmath,graphicx}

\usepackage{amsmath}
\usepackage{flexisym}
\usepackage{dcolumn}
\usepackage{kotex}
\usepackage{color}
\usepackage{hhline}
\usepackage{verbatim}
\usepackage{multirow}
\usepackage{amssymb}
\usepackage{rotating}
\usepackage{subfigure}
\usepackage{dblfloatfix}
\usepackage{enumitem}
\newcolumntype{L}[1]{>{\raggedright\let\newline\\\arraybackslash\hspace{0pt}}m{#1}}
\newcolumntype{C}[1]{>{\centering\let\newline\\\arraybackslash\hspace{0pt}}m{#1}}
\newcolumntype{R}[1]{>{\raggedleft\let\newline\\\arraybackslash\hspace{0pt}}m{#1}}

\newcommand\Tstrut{\rule{0pt}{2.5ex}}         

\title{Speech Emotion Recognition Using Multi-Hop Attention Mechanism}
\name{Seunghyun Yoon$^{1,3}$\sthanks{Most work was done while the author was an intern at Adobe Research.}, Seokhyun Byun$^{1}$, Subhadeep Dey$^{2,3}$ and Kyomin Jung$^{1}$}
\address{
$^{1}$Department of Electrical and Computer Engineering, \\ Seoul National University, Seoul, Korea \\
$^{2}$Idiap Research Institute, Martigny, Switzerland \\
$^{3}$Adobe Research, San Jose, USA \\
{
mysmilesh@snu.ac.kr,~byuns9334@snu.ac.kr,~subhadeep.dey@idiap.ch,~kjung@snu.ac.kr
}
}

\begin{document}
\ninept
\maketitle
\begin{abstract}

In this paper, we are interested in exploiting textual and acoustic data of an utterance for the speech emotion classification task. The baseline approach models the information from audio and text independently using two deep neural networks (DNNs). The outputs from both the DNNs are then fused for classification. As opposed to using knowledge from both the modalities separately, we propose a framework to exploit acoustic information in tandem with lexical data. The proposed framework uses two bi-directional long short-term memory (BLSTM) for obtaining hidden representations of the utterance. Furthermore, we propose an attention mechanism, referred to as the multi-hop, which is trained to automatically infer the correlation between the modalities. 
The multi-hop attention first computes the relevant segments of the textual data corresponding to the audio signal. The relevant textual data is then applied to attend parts of the audio signal. To evaluate the performance of the proposed system, experiments are performed in the IEMOCAP dataset. Experimental results show that the proposed technique outperforms the state-of-the-art system by 6.5\% relative improvement in terms of weighted accuracy.

\end{abstract}
\section{Introduction}
\label{sec:intro}

In this era of high-performance computing, human-computer interaction (HCI) has become pervasive.
To enrich the user experience, the system is often required to detect human emotion and produce a response with proper emotional context~\cite{picard2003affective, busso2013toward}. The first step in such an HCI involves building a system that recognizes emotion from the speech utterance. A speech emotional system aims to identify audio recording as belonging to one of the categories, like happy, sad, angry or neutral. Beside HCI, the output of emotion recognition engine is beneficial in the paralinguistic area as well~\cite{kolakowska2014emotion}.
In this paper, we build a speech emotion recognition system that uses acoustic and textual information in tandem.

Various approaches to address emotion recognition have been investigated in the literature. Most of the techniques involve extracting low-level or high-level acoustic features for this task~\cite{han2014speech}. In emotion recognition, the lexical content of the audio recording is an important source of information that is usually ignored. For example, the presence of words such as ``gorgeous" and ``stunning" in the utterance would indicate that the person is happy. Recently researchers have also explored the application of textual content of the speech signal for this task.
In~\cite{schuller2004speech}, frame and supra-segmental level features (such as pitch and spectral contours) are derived from the speech signal. Textual information is used by spotting keywords that emphases the emotional states of the speaker. The work in~\cite{jin2015speech} also presents an approach to exploit the acoustic and lexical content. 
In particular, they explored conventional acoustic features from the speech signal while the textual information is derived from the bag of word representation.

Recently, deep neural network (DNN) has shown to provide good results for modeling acoustic and textual information for emotion identification. 
In~\cite{schuller2004speech}, textual and acoustic information of the utterance are used by a DNN to obtain hidden feature representations for both the modality. These features are then concatenated to represent the utterance and subsequently used to classify the emotion of the speaker. Experimental evidence shows the potential of the approach. In our previous work~\cite{yoon2018multimodal}, we applied a dual RNN in order to obtain a richer representation by blending the content and acoustic knowledge.

In this paper, we improve upon our earlier work by incorporating an attention mechanism in the emotion recognition framework. The proposed attention mechanism is trained to exploit both textual and acoustic information in tandem. We refer to this attention method as the multi-hop. The multi-hop attention is designed to select relevant parts of the textual data, which is subsequently applied to attend to the segments of the audio signal for classification. We hypothesize that this approach would automatically detect the segments that contain information relevant for the task. The emotion recognition experiments are performed on the standard IEMOCAP dataset~\cite{busso2008iemocap}. Experimental results indicate that the proposed approach outperforms the state-of-the-art system published in the literature on this database by 6.5\% relative improvement in terms of weighted accuracy.

This paper is organized as follows. Section~\ref{sec:related} provides a brief literature review on speech emotion recognition. In Section~\ref{sec:model}, we start by describing the baseline bidirectional recurrent encoder model considered in this paper, then introducing the proposed technique in detail.
Experimental setup for evaluating the system and discussion of the achieved results by various systems are presented in Sections~\ref{sec:experiments}.
Finally, the paper is concluded in Section~\ref{sec:conclusions}.

\section{Related work}
\label{sec:related}

Along with classical algorithms based models such as support vector machine (SVM), hidden markov model (HMM) and decision tree~\cite{seehapoch2013speech, schuller2003hidden, lee2011emotion}, various neural network architectures have been recently introduced for the speech emotion recognition task.
For example, convolutional neural network (CNN)-based models were trained on spectrograms or audio features such as mel-frequency cepstral coefficients (MFCCs) and low-level descriptors (LLDs)~\cite{bertero2017first, badshah2017speech, aldeneh2017using}. 
More complex models such as~\cite{satt2017efficient} were designed to better learn nonlinear decision boundaries of emotional speech and achieved the best-recorded performance in audio modality models on IEMOCAP dataset~\cite{busso2008iemocap}. Several neural network models with attention mechanism have been proposed to efficiently focus on a prominent part of speech and learn temporal dependency within whole utterance~\cite{li2018attention,mirsamadi2017automatic}.

\input{rsc/fig_MHA.tex}

Multi-modal approaches using acoustic features and textual information have been investigated.
\cite{schuller2004speech} identified emotional key phrases and salience of verbal cues from both phoneme sequences and words. Recently,~\cite{yoon2018multimodal,cho2018deep} combined acoustic information and conversation transcripts using a neural network-based model to improve emotion classification accuracy.
However, none of these studies utilized attention method over audio and text modality in tandem for contextual understanding of the emotion in audio recording.

\section{Model}
\label{sec:model}
This section describes the methodologies that are applied to the speech emotion recognition task. We start by introducing a baseline model, the bidirectional recurrent encoder, for encoding the audio and text modalities individually. 
We then propose an approach to exploit both audio and text data in tandem. In this technique, multi-hop attention is proposed to obtain relevant parts of audio and text data automatically.

\subsection{Bidirectional Recurrent Encoder}
Motivated by the architecture used in~\cite{yoon2018multimodal, mirsamadi2017automatic, wang2016audio}, we train a recurrent encoder to predict the categorical class of a given audio signal. To model the sequential nature of the speech signal, we use a bi-directional recurrent encoder (\textbf{BRE}) as shown in the Figure~\ref{fig_BRE}. We also added a residual connection to the model for promoting convergence during training~\cite{wang2016recurrent}. 
A sequence of feature vectors is fed as input to the \textbf{BRE}, which leads to the formation of hidden states of the model as given by the following equation:  

\begin{equation}
\begin{aligned}
& \overrightarrow{\textbf{h}}_{t} = f_{\theta}( \overrightarrow{\textbf{h}}_{t-1}, \overrightarrow{\textbf{x}}_{t}) +\overrightarrow{\textbf{x}_t},\\
& \overleftarrow{\textbf{h}}_{t} = f'_{\theta}( \overleftarrow{\textbf{h}}_{t+1}, \overleftarrow{\textbf{x}}_{t}) +\overleftarrow{\textbf{x}_t},\\
& \textbf{o}_{t} = [\overrightarrow{\textbf{h}}_{t} ;\overleftarrow{\textbf{h}}_{t}],\\
&\textbf{o}^{A}_{t}=[\textbf{o}_{t} ;\textbf{p}],
\end{aligned}
\label{eq_BRE}
\end{equation}
where $f_{\theta}$, $f'_{\theta}$ are the forward and backward long short-term memory (LSTM)   with weight parameter $\theta$, $\textbf{h}_t$ represents the hidden state at \textit{t}-th time step, and $\textbf{x}_t$ represents the \textit{t}-th MFCC features in audio signal. The hidden representations ($\overrightarrow{\textbf{h}}_{t}$, $\overleftarrow{\textbf{h}}_{t}$) from forward/backward LSTMs are concatenated for produce the feature, $\mathbf{o}_t$.
To follow previous research~\cite{yoon2018multimodal}, we also add another prosodic feature vector, $\textbf{p}$, with each $\textbf{o}_{t}$ to generate a more informative vector representation of the signal, $\textbf{o}^{A}_{t}$.
Finally, an emotion class is predicted from the acoustic signal by applying a softmax function to the final hidden representation at the last time step,  $\textbf{o}^{A}_{\text{last}}$.
We refer this model as \textbf{audio-BRE} with the objective function as follows:
\begin{equation}
\begin{aligned}
\hat{y}_{c} = \text{softmax}(~({\textbf{o}^{A}_{\text{last}})}^\intercal~\textbf{W}+\textbf{b}~), \\
\mathcal{L} = -\log \prod_{i=1}^{N} \sum_{c=1}^{C} y_{i,c} \text{log} (\hat{y}_{i,c}),
\end{aligned}
\label{eq_BRE_loss}
\end{equation}
where $y_{i,c}$ is the true label vector, and $\hat{y}_{i,c}$ is the predicted probability distribution from the softmax layer.
The $\textbf{W}$ and the bias $\textbf{b}$ are learned model parameters. $C$ is the total number of classes, and $N$ is the total number of samples used in training.

Next, we attempt to use the processed textual information as another modality in predicting the emotion class of a given signal.
To obtain textual hidden representation, $\textbf{o}_{t}^{T}$, we tokenize the transcript and feed it into the \textbf{BRE} in such a way that the acoustic signals are encoded by equation~(\ref{eq_BRE}).
We refer this model as \textbf{text-BRE}.
The training objective for the \textbf{text-BRE} is same as the \textbf{audio-BRE} in equation~(\ref{eq_BRE_loss}).


\subsection{Proposed Multi-Hop Attention}
We propose a novel multi-hop attention method to predict the importance of audio and text, referred to multi-hop attention (\textbf{MHA}).
Figure~\ref{fig_MHA} shows the architecture of the proposed \textbf{MHA} model. Previous research used multi-modal information independently using neural network model by concatenating features from each modality~\cite{yoon2018multimodal,tripathi2018multi}. As opposed to this approach, we propose a neural network architecture that exploits information in each modality by extracting relevant segments of the speech data using information from the lexical content (and vice-versa). 

First, the acoustic and textual data are encoded with the \textbf{audio-BRE} and \textbf{text-BRE}, respectively, using equation~(\ref{eq_BRE}).
We then consider the final hidden representation of \textbf{audio-BRE}, $\textbf{o}^{A}_{\text{last}}$, as a context vector and apply attention method to the textual sequence, $\textbf{o}_{t}^{T}$. As this model is developed with a single attention method, we refer to the model as \textbf{MHA-1}. The final hidden representation of the \textbf{MHA-1} model, $\textbf{\text{H}}$, is calculated as follows:
\begin{equation}
\begin{aligned}
	&a_i=\dfrac{\text{exp}({~(\textbf{o}^{A}_{\text{last}})}^\intercal~\textbf{o}^{T}_{i}~)}{\sum_{i} \text{exp}({~(\textbf{o}^{A}_{\text{last}})}^\intercal~\textbf{o}^{T}_{i}~)},~(i=1,...,t)
    \\ 
    &~\textbf{H}^1={\sum_{i}} a_{i}~{\textbf{o}^{T}_{i}},
    ~~\textbf{H}=[\textbf{H}^1 ;\textbf{o}^{A}_{\text{last}}].
\end{aligned}
\label{eq_hot_1}
\end{equation}

The $\textbf{H}^1$ (equation~\ref{eq_hot_1}) is a new hidden representation for textual information with consideration of audio modality. With this information, we apply 2nd-hop attention, referred to \textbf{MHA-2}, to the audio sequence. The final hidden representation of the \textbf{MHA-2} model, \textbf{H}, is calculated as follows:
\begin{equation}
\begin{aligned}
	&a_{i}=\dfrac{~\text{exp}({~(\text{\textbf{H}}_1)}^\intercal~\textbf{o}^{A}_{i}~)}{\sum_i \text{exp}({~(\text{\textbf{H}}_1)}^\intercal~\textbf{o}^{A}_{i}~)},~(i=1,...,t)
    \\ 
   & ~\textbf{H}^2=\sum_{i} a_{i}~{\textbf{o}^{A}_{i}},
    ~~\textbf{H}=[\textbf{H}^1 ;\textbf{H}^2],
\end{aligned}
\label{eq_MHA_2}
\end{equation}
where $\textbf{H}^2$ is a new hidden representation for audio information with the consideration of textual modality obtained from the \textbf{MHA-1}.

Similarly to \textbf{MHA-1}, we further apply 3rd-hop attention to textual sequence, referred to \textbf{MHA-3}, with the new audio hidden representation $\textbf{H}^2$ (equation~\ref{eq_MHA_2}). The final hidden representation of the \textbf{MHA-3} model, \textbf{H}, is calculated as follows:
\begin{equation}
\begin{aligned}
	&a_{i}=\dfrac{\text{exp}(~({\text{\textbf{H}}_2})^\intercal~\textbf{o}^{T}_{i})}{\sum_{i} \text{exp}(~({\text{\textbf{H}}_2})^\intercal~\textbf{o}^{T}_{i})},~(i=1,...,t)
    \\ 
   & ~\textbf{H}^3=\sum_{i} a_i~{\textbf{o}^{T}_{i}},
    ~~\textbf{H}=[\textbf{H}^3 ;\textbf{H}^2],
\end{aligned}
\label{eq_MHA_3}
\end{equation}
where $\textbf{H}^3$ is updated representative vector of the textual information with the consideration of audio modality one more time.

In each case, the final hidden representation, $\textbf{H}$, is passed through the softmax function to predict the four-categories emotion class.
We use the same training objective as the \textbf{BRE} model with equation~(\ref{eq_BRE_loss}), and the predicted probability distribution for the target class, $\hat{y}_{c}$ is as follows:
\begin{equation}
\begin{aligned}
   &\hat{y}_{c} = \text{softmax}((\textbf{H})^\intercal~\textbf{W}+\textbf{b}~), \\
\end{aligned}
\label{eq_MHA_loss}
\end{equation}
where projection matrix $\textbf{W}$ and bias $\textbf{b}$ are leaned model parameters.


\section{Experiments}
\label{sec:experiments}
\subsection{Dataset and Experimental Setup}
To train and evaluate our model, we use the Interactive Emotional Dyadic Motion Capture (IEMOCAP)~\cite{busso2008iemocap} dataset, which includes five sessions of utterances between two speakers (one male and one female). Total 10 unique speakers participated in this work.
For consistent comparison with previous works~\cite{yoon2018multimodal,cho2018deep}, all utterances labeled ``excitement" are merged with those labeled ``happiness". We assign single categorical emotion to the utterance with majority of annotators agreed on the emotion labels. The final dataset contains 5,531 utterances  in total (1,636 \textit{happy}, 1,084 \textit{sad}, 1,103 \textit{angry} and 1,708 \textit{neutral}).
In the training process, we perform 10-fold cross-validation where each 8, 1, 1 folds are used for the train set, development set, and test set, respectively.

\subsection{Feature extraction and Implementation details}
As this research is extended work from previous research~\cite{yoon2018multimodal}, we use the same feature extraction method as done in our previous work. 
After extracting 40-dimensional Mel-frequency cepstral coefficients (MFCC) feature (frame size is set to 25 ms at a rate of 10 ms with the Hamming window) using Kaldi~\cite{povey2011kaldi}, we concatenate it with its first, second order derivates, making the feature dimension to 120. We also extract prosodic features by using OpenSMILE toolkit~\cite{eyben2013recent} and appending it to the audio feature vector.

In preparing the textual dataset, we first use the ground-truth transcripts of the IEMOCAP dataset.
In a practical scenario where we may not access to transcripts of the audio, we obtain all of the transcripts from the speech signal using a commercial ASR system~\cite{GoogleCloudSpeechAPI} (The performance of the ASR system is word error rate (WER) of 5.53\%). We apply word-tokenizer to the transcripts and obtain sequential data for textual input.

The maximum length of an audio segment is set to 750 based on the implementation choices presented in~\cite{neumann2017attentive} and 128 for the textual input which covers the maximum length of the tokenized transcripts.
We minimize the cross-entropy loss function using (equation~(\ref{eq_BRE_loss})) the Adam optimizer~\cite{kingma2014adam} with a learning rate of 1e-3 and gradients clipped with a norm value of 1. For the purposes of regularization, we apply the dropout method, 30\%. 
The number of hidden units and the number of layers in the RNN for each model (\textbf{BRE} and \textbf{MHA}) are optimized on the development set.




\subsection{Performance evaluation}

To measure the performance of systems, we report the weighted accuracy (WA) and unweighted accuracy (UA) averaging over the 10-fold cross-validation experiments.
We use the same dataset and features as other researchers~\cite{yoon2018multimodal,cho2018deep}.


Table~\ref{table_performance} presents performances of proposed approaches for recognizing speech emotion in comparison with various models.
To compare our results from previous approaches, we first use ground-truth transcripts included in the dataset in training textual modality.
From the previous model, \textbf{E\_vec-MCNN-LSTM} encodes acoustic signal and textual information using a neural network (RNN and CNN, respectively) and then fuse each result by concatenating and feeding them into following (SVM) to predict emotion labels. On the other hand, \textbf{MDRE} model use dual-RNNs to encode both the modalities and merge the results using another fully-connect neural network layer. This \textbf{MDRE} approach applies end-to-end learning and outperforms \textbf{E\_vec-MCNN-LSTM} by 10.6\% relative (0.649 to 0.718 absolute) in terms of WA.

\input{rsc/table_performance.tex}

\input{rsc/fig_confusion.tex}

Among our proposed system, the \textbf{audio-BRE} model that uses an acoustic signal with bidirectional-RNN architecture achieves WA 0.646. Interestingly, the \textbf{text-BRE} model that use textual information shows higher performance than that of \textbf{audio-BRE} by 8\% relative (0.646 to 0.698) in WA.
The multi-hop attention model, \textbf{MHA-N}, ($N$ = 1, 2, 3), shows a substantial performance gain. In particular, the \textbf{MHA-2} model (best performing system among MHA-N) outperformed the best baseline model, \textbf{MDRE}, by 6.5\% relative (0.718 to 0.765) in WA.
Although we observe performance degradation in the \textbf{MHA-3} model, we believe that this could be due to the limited data for training.

In a practical scenario, we may not access the audio transcripts. 
We describe the effect of using ASR-processed transcripts on the proposed system. 
From table~\ref{table_performance}, we observe performance degradation in \textbf{text-BRE-ASR} and \textbf{MHA-2-ASR} (our best system), compared to that of \textbf{text-BRE} and \textbf{MHA-2} by 6.6\% (0.698 to 0.652) and 4.6\% (0.765 to 0.730) relative in WA, receptively.
Even with the erroneous transcripts (WER = 5.53\%), however, the proposed approach (\textbf{MHA-2-ASR}) outperforms the best baseline system (\textbf{MDRE}) by 1.6\% relative (0.718 to 0.730) in terms of WA.


\subsection{Error analysis}

Figure~\ref{fig_confusion} shows the confusion matrices of the proposed systems.
In \textbf{audio-BRE} (Fig.~\ref{fig_confusion_audio}), most of the emotion labels are frequently misclassified as \textit{neutral} class, supporting the claims of~\cite{yoon2018multimodal,neumann2017attentive}.
The \textbf{text-BRE} shows improvement in classifying most of the labels in Fig.~\ref{fig_confusion_text}. 
In particular, \textit{angry} and \textit{happy} classes are correctly classified by 32\% (57.14 to 75.41) and 63\% (40.21 to 65.56) relative in accuracy with respect to \textbf{audio-BRE}, receptively.
However, it incorrectly predicted instances of the \textit{happy} class as \textit{sad} class in 10\% of the time, even though these emotional states are opposites of one another.

The \textbf{MHA-2} (our best system, Fig.~\ref{fig_confusion_MHA}) compensates for the weaknesses of the single modality models and benefits from their strengths. It shows significant performance gain for 
\textit{angry}, \textit{happy}, \textit{sad} and \textit{neutral} classes by 6\%, 20\%, 15\% and 13\% relative in accuracy with respect to \textbf{text-BRE}. It also correctly classify \textit{neutral} class similar to that of \textbf{audio-BRE} (81.63 and 78.00 for \textbf{audio-BRE} and \textbf{MHA-2}, receptively).
Interestingly, although \textbf{MHA-2} shows superior discriminating ability among emotion classes, it still shows the tendency such that most of the incorrect cases are misclassified into \textit{neutral} class. We consider this observation as a future research direction.

\section{Conclusions}
\label{sec:conclusions}
In this paper, we propose a multi-hop attention model to combine acoustic and textual data for speech emotion recognition task. 
The proposed attention method is designed to select relevant parts of the textual data, which is subsequently applied to attend to the segments of the audio signal for classification.
Extensive experiments show that the proposed \textbf{MHA-2} outperforms the best baseline system in classifying the four emotion categories by 6.5\% (0.718 to 0.765 absolute) in terms of WA when the model is applied to the IEMOCAP dataset.
We further test our model with ASR-processed transcripts and achieve WA 0.73 that shows the reliability of the proposed system (\textbf{MHA-2-ASR}) in the practical scenario where the ground-truth transcripts are not available.

\section*{Acknowledgments}
We sincerely thank Trung H. Bui at Adobe Research for his in depth feedback that helped us to think the technology from the industry point of view as well.
K. Jung and S. Yoon are with Automation and Systems Research Institute (ASRI), Seoul National University, Seoul, Korea. This work was supported by the Ministry of Trade, Industry \& Energy (MOTIE, Korea) under Industrial Technology Innovation Program (No.10073144) and  by the National Research Foundation of Korea (NRF) funded by the Korea government (MSIT) (No. 2016M3C4A7952632).


\bibliographystyle{IEEEbib}
\bibliography{ICASSP19}


\end{document}

%% file: rsc/fig_MHA.tex
\begin{figure*}[!t]
\small
\centering
\subfigure[BRE]{\label{fig_BRE}\includegraphics[width=0.4\columnwidth]{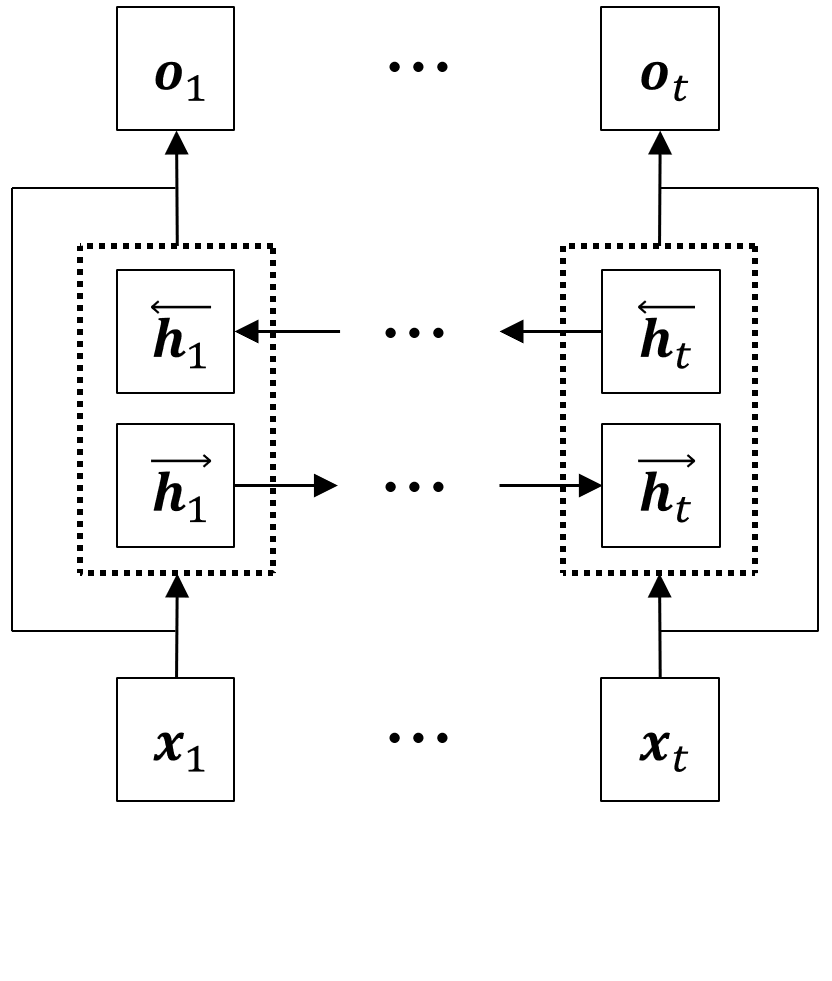}}
\quad
\qquad
\subfigure[MHA-1]{\label{fig_MHA_1}\includegraphics[width=0.42\columnwidth]{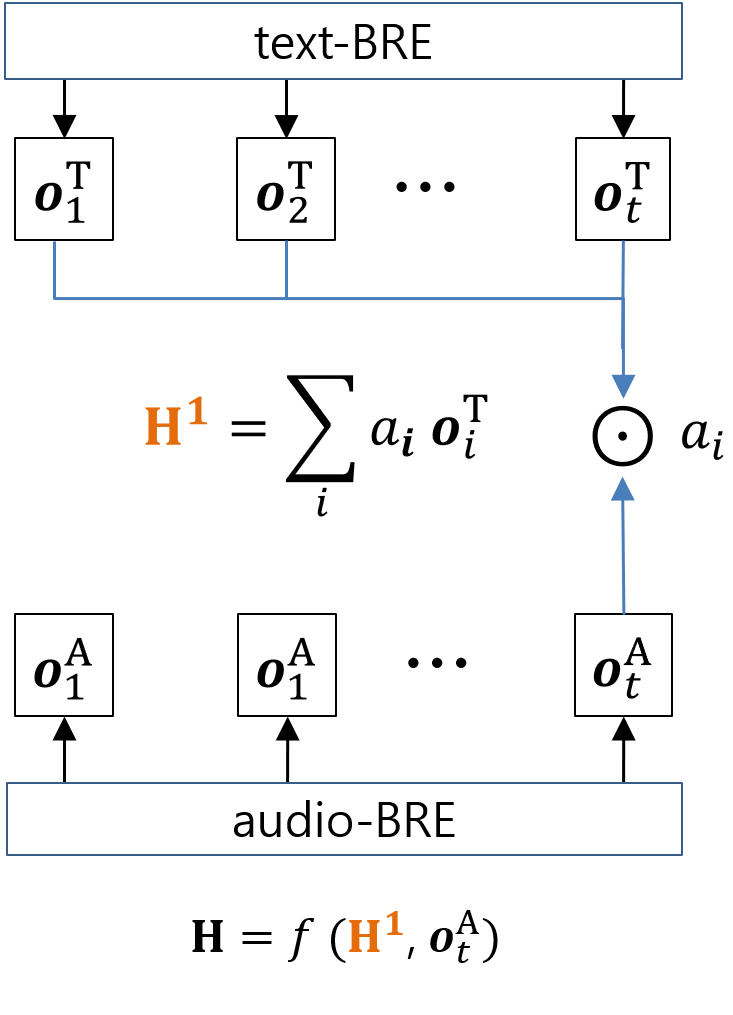}}
\quad
\qquad
\subfigure[MHA-2]{\label{fig_MHA_2}\includegraphics[width=0.42\columnwidth]{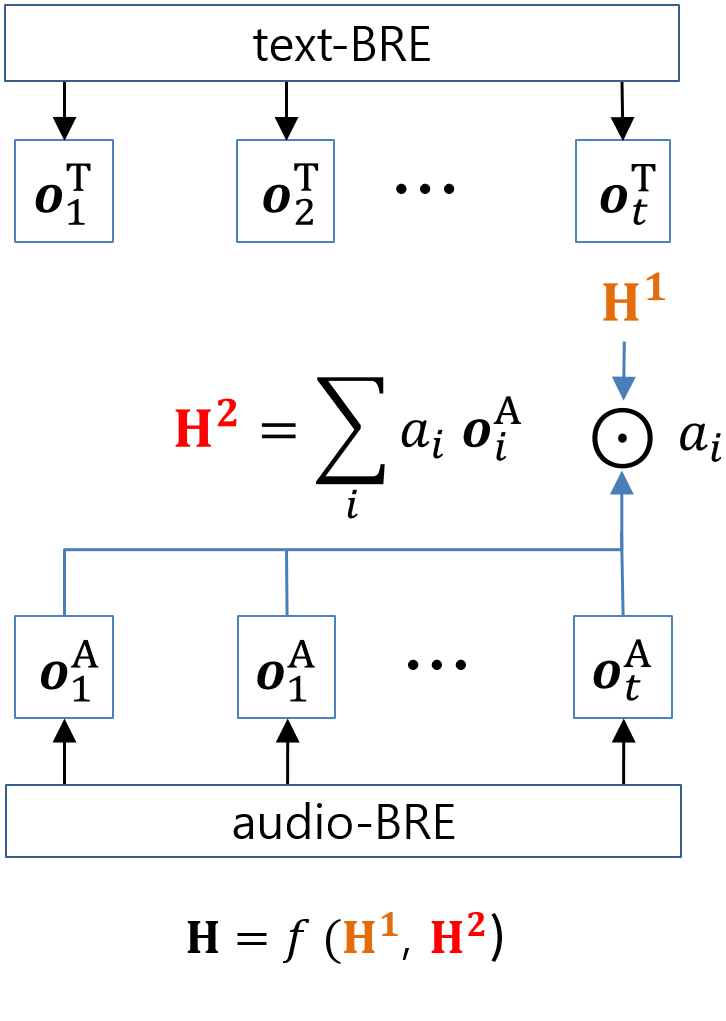}}
\quad
\qquad
\subfigure[MHA-3]{\label{fig_MHA_3}\includegraphics[width=0.42\columnwidth]{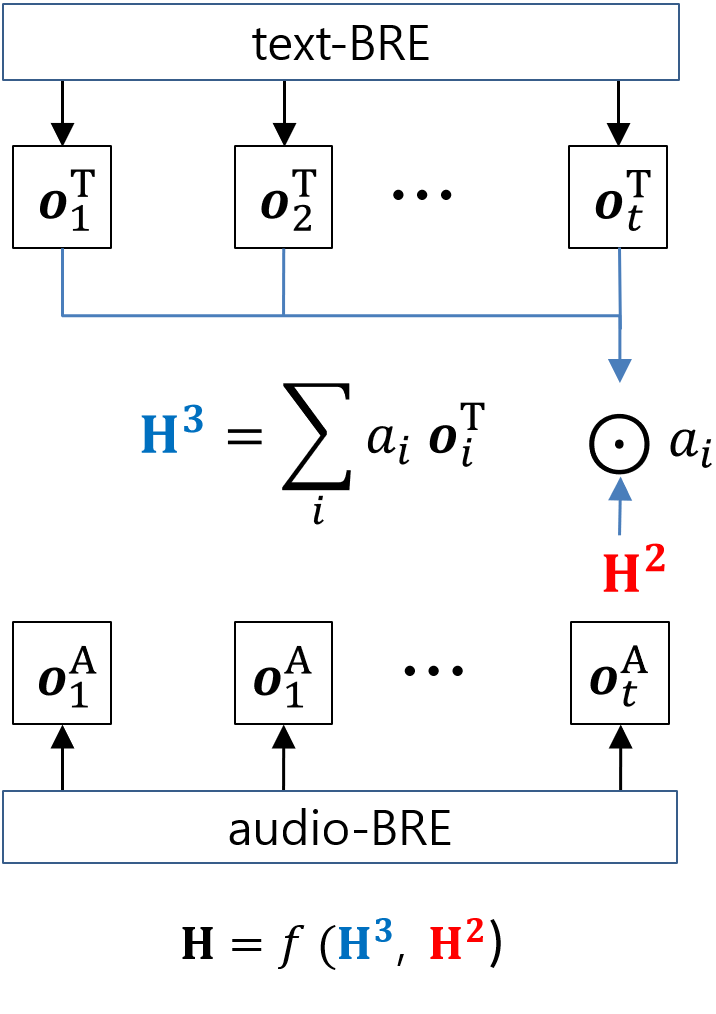}}
\caption{
Architecture of the bidirectional recurrent encoder (BRE), and multi-hop attention model (MHA).
}
\label{fig_MHA}
\end{figure*}

%% file: rsc/table_performance.tex
\begin{table}[!t]
\small
\centering
\begin{tabular}
{C{0.45\columnwidth}C{0.15\columnwidth}C{0.10\columnwidth}C{0.10\columnwidth}}
\hline 
\textbf{Model}\Tstrut & \textbf{Modality} & \textbf{WA}  & \textbf{UA} \\ 
\hline
\multicolumn{4}{c}{Ground-truth transcript} \\
\hline
\textbf{E\_vec-MCNN-LSTM}~\cite{cho2018deep}\Tstrut 	       & A+T\Tstrut	& 0.649\Tstrut & 0.659\Tstrut   \\
\textbf{MDRE}~\cite{yoon2018multimodal} 	   & A+T	& 0.718 & -   \\

\hline
\textbf{audio-BRE} (ours)\Tstrut      & A\Tstrut	  & 0.646\Tstrut  & 0.652\Tstrut  \\
\textbf{text-BRE} (ours)        & T   & 0.698  & 0.703  \\
\textbf{MHA-1} (ours)       & A+T & \textbf{0.756}  & \textbf{0.765}  \\
\textbf{MHA-2} (ours)       & A+T & \textbf{0.765}  & \textbf{0.776}  \\
\textbf{MHA-3} (ours)       & A+T & 0.740  & 0.753  \\
\hline
\multicolumn{4}{c}{ASR-processed transcript} \\
\hline
\textbf{text-BRE-ASR} (ours)\Tstrut & T\Tstrut    & 0.652\Tstrut  & 0.658\Tstrut  \\
\textbf{MHA-2-ASR} (ours)       & A+T  & 0.730  & 0.739  \\

\hline 

\end{tabular}
\caption{
  Model performance comparisons. The top 2 best-performing models (according to the WA) are marked in bold. The ``-ASR'' models use ASR-processed transcripts from the Google Cloud Speech API. The ``A'' and ``T'' in modality indicate ``Audio'' and ``Text'', receptively.
}
\label{table_performance}
\end{table}

%% file: rsc/fig_confusion.tex
\begin{figure*}[!b]
\small
\centering
\subfigure[audio-BRE]{\label{fig_confusion_audio}\includegraphics[width=0.60\columnwidth]{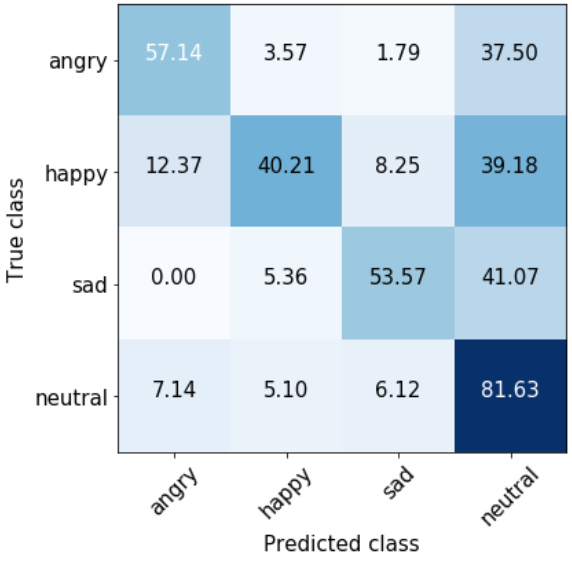}}
\qquad
\subfigure[text-BRE]{\label{fig_confusion_text}\includegraphics[width=0.60\columnwidth]{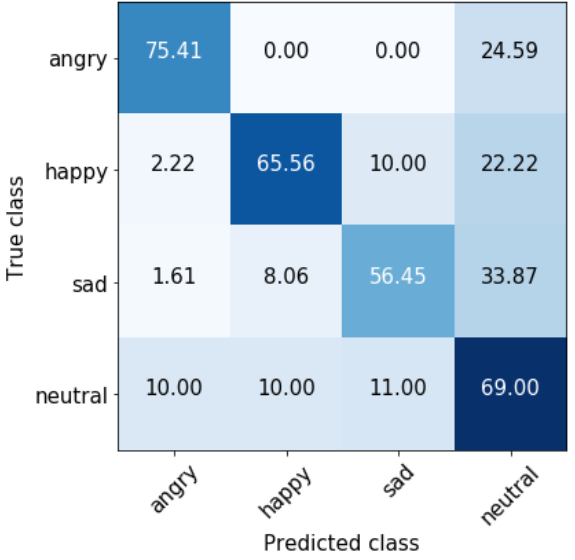}}
\qquad
\subfigure[MHA-2]{\label{fig_confusion_MHA}\includegraphics[width=0.60\columnwidth]{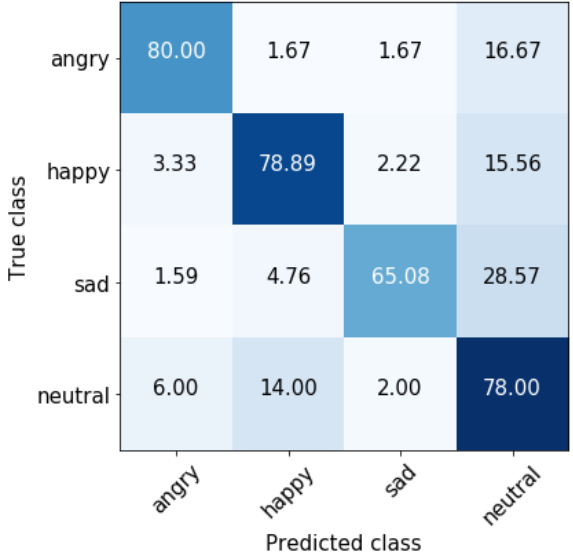}}
\caption{
Confusion matrix of each model.
}
\label{fig_confusion}
\end{figure*}